%% file: main.tex
\pdfoutput=1

\documentclass[sigconf,nonacm]{acmart}

\usepackage{pvldb}

\renewcommand\vldbdoi{XX.XX/XXX.XX}
\renewcommand\vldbpages{XXX-XXX}
\renewcommand\vldbavailabilityurl{}

\usepackage{amsmath,amsthm}
\makeatletter
\@ifpackageloaded{newtxmath}{}{\usepackage{amssymb}}
\makeatother
\usepackage{graphicx}
\usepackage{booktabs,multirow}
\usepackage{microtype}
\usepackage{xcolor}
\usepackage{pgfplots}
\pgfplotsset{compat=1.18}
\usepgfplotslibrary{groupplots}
\usepackage{colortbl}
\usepackage{hyperref}
\usepackage{cleveref}
\usepackage[normalem]{ulem}
\usepackage{algorithm}
\usepackage{algpseudocode}
\usepackage{listings}
\usepackage{tikz}
\usetikzlibrary{positioning,arrows.meta,fit,calc}

\lstdefinestyle{prompt}{%
  basicstyle=\ttfamily\footnotesize,
  breaklines=true,
  breakindent=0pt,
  columns=fullflexible,
  keepspaces=true,
  frame=single,
  framesep=4pt,
  xleftmargin=5pt,
  xrightmargin=5pt,
  aboveskip=5pt,
  belowskip=5pt,
}
\hypersetup{
  colorlinks=true,
  linkcolor=blue!50!black,
  citecolor=green!50!black,
  urlcolor=blue!50!black,
}

\title{MetaJoin: Adaptive Join Selections for Heterogeneous Tables}

\author{Ning Wang}
\affiliation{%
  \institution{Cornell University}
  \country{} 
}
\email{nw366@cornell.edu}

\author{Sainyam Galhotra}
\affiliation{%
  \institution{Cornell University}
  \country{} 
}
\email{sg@cs.cornell.edu}

\RequirePackage[normalem]{ulem}
\RequirePackage{color}

\definecolor{RED}{rgb}{1,0,0}
\definecolor{BLUE}{rgb}{0,0,1}

\providecommand{\DIFaddbegin}{}
\providecommand{\DIFaddend}{}
\providecommand{\DIFdelbegin}{}
\providecommand{\DIFdelend}{}

\providecommand{\DIFaddbeginFL}{}
\providecommand{\DIFaddendFL}{}
\providecommand{\DIFdelbeginFL}{}
\providecommand{\DIFdelendFL}{}

\RequirePackage{settobox}
\RequirePackage{letltxmacro}

\newsavebox{\DIFdelgraphicsbox}
\newlength{\DIFdelgraphicswidth}
\newlength{\DIFdelgraphicsheight}

\LetLtxMacro{\DIFOincludegraphics}{\includegraphics}

\providecommand{\DIFaddincludegraphics}[2][]{%
  {\color{blue}\fbox{\DIFOincludegraphics[#1]{#2}}}%
}

\providecommand{\DIFdelincludegraphics}[2][]{%
  \sbox{\DIFdelgraphicsbox}{\DIFOincludegraphics[#1]{#2}}%
  \settoboxwidth{\DIFdelgraphicswidth}{\DIFdelgraphicsbox}%
  \settoboxtotalheight{\DIFdelgraphicsheight}{\DIFdelgraphicsbox}%
  \scalebox{\DIFscaledelfig}{%
    \parbox[b]{\DIFdelgraphicswidth}{%
      \usebox{\DIFdelgraphicsbox}\\[-\baselineskip]
      \rule{\DIFdelgraphicswidth}{0em}%
    }%
    \llap{%
      \resizebox{\DIFdelgraphicswidth}{\DIFdelgraphicsheight}{%
        \setlength{\unitlength}{\DIFdelgraphicswidth}%
        \begin{picture}(1,1)
          \thicklines
          \linethickness{2pt}
          {\color[rgb]{1,0,0}\put(0,0){\framebox(1,1){}}}
          {\color[rgb]{1,0,0}\put(0,0){\line(1,1){1}}}
          {\color[rgb]{1,0,0}\put(0,1){\line(1,-1){1}}}
        \end{picture}%
      }%
      \hspace*{3pt}%
    }%
  }%
}

\LetLtxMacro{\DIFOaddbegin}{\DIFaddbegin}
\LetLtxMacro{\DIFOaddend}{\DIFaddend}
\LetLtxMacro{\DIFOdelbegin}{\DIFdelbegin}
\LetLtxMacro{\DIFOdelend}{\DIFdelend}

\DeclareRobustCommand{\DIFaddbegin}{%
  \DIFOaddbegin
  \let\includegraphics\DIFaddincludegraphics
}
\DeclareRobustCommand{\DIFaddend}{%
  \DIFOaddend
  \let\includegraphics\DIFOincludegraphics
}
\DeclareRobustCommand{\DIFdelbegin}{%
  \DIFOdelbegin
  \let\includegraphics\DIFdelincludegraphics
}
\DeclareRobustCommand{\DIFdelend}{%
  \DIFOdelend
  \let\includegraphics\DIFOincludegraphics
}

\LetLtxMacro{\DIFOaddbeginFL}{\DIFaddbeginFL}
\LetLtxMacro{\DIFOaddendFL}{\DIFaddendFL}
\LetLtxMacro{\DIFOdelbeginFL}{\DIFdelbeginFL}
\LetLtxMacro{\DIFOdelendFL}{\DIFdelendFL}

\DeclareRobustCommand{\DIFaddbeginFL}{%
  \DIFOaddbeginFL
  \let\includegraphics\DIFaddincludegraphics
}
\DeclareRobustCommand{\DIFaddendFL}{%
  \DIFOaddendFL
  \let\includegraphics\DIFOincludegraphics
}
\DeclareRobustCommand{\DIFdelbeginFL}{%
  \DIFOdelbeginFL
  \let\includegraphics\DIFdelincludegraphics
}
\DeclareRobustCommand{\DIFdelendFL}{%
  \DIFOdelendFL
  \let\includegraphics\DIFOincludegraphics
}

\makeatletter
\let\sout@orig\sout
\renewcommand{\sout}[1]{%
  \ifmmode
    \text{\sout@orig{\ensuremath{#1}}}%
  \else
    \sout@orig{#1}%
  \fi
}
\makeatother

\RequirePackage{listings}
\RequirePackage{color}

\lstdefinelanguage{DIFcode}{%
  moredelim=[il][\color{red}\sout]{\%DIF\ <\ },
  moredelim=[il][\color{blue}\uwave]{\%DIF\ >\ }
}

\lstdefinestyle{DIFverbatimstyle}{%
  language=DIFcode,
  basicstyle=\ttfamily,
  columns=fullflexible,
  keepspaces=true
}

\lstnewenvironment{DIFverbatim}[1][]{%
  \lstset{style=DIFverbatimstyle}%
}{}

\lstnewenvironment{DIFverbatim*}[1][]{%
  \lstset{style=DIFverbatimstyle,showspaces=true}%
}{}

\makeatletter
\@ifundefined{DIFOincludegraphics}{}{%
  \let\DIFaddincludegraphics\DIFOincludegraphics
  \let\DIFdelincludegraphics\DIFOincludegraphics
}
\makeatother
\begin{document}

\begin{abstract}
Joining heterogeneous tables requires different forms of correspondence, including exact agreement, syntactic transformation, semantic similarity, and entity-level matching. The appropriate join method is often unknown in advance, while running every candidate method incurs unnecessary cost. We introduce \textsc{MetaJoin}, a label-free framework for \emph{adaptive join selection}. MetaJoin uses input evidence to choose an initial specialist, then uses specialist-specific output evidence to determine whether that result should be retained, compared with an alternative, escalated, or repaired. This allows join selection to adapt after execution rather than committing to a one-shot routing decision. We also introduce \textsc{MNN Join}, a transformation specialist that searches for string transformations by maximizing mutual-nearest-neighbor coverage between two unaligned columns and uses the same reciprocal structure to validate its output. We evaluate \textsc{MetaJoin} on 670 datasets from six benchmarks. It achieves $0.817$ macro row-$F_1$, slightly exceeding a benchmark-level oracle that is given the best standalone specialist for each benchmark ($0.812$), and outperforms label-free selection policies built from the same specialist portfolio by more than $0.13$. Across the 609 datasets in the four transformation-dominant benchmarks, \textsc{MNN Join} outperforms every evaluated transformation baseline. Finally, \textsc{MetaJoin}'s accuracy varies by only $0.015$ as the controller backbone ranges from $3$B to $14$B parameters, showing that output validation and repair substantially reduce dependence on the initial routing decision.
\end{abstract}
\maketitle


\input{intro_new}
\input{related}
\input{method}
\input{exp_new}
\input{discussion}    

\bibliographystyle{ACM-Reference-Format}
\bibliography{references}


\end{document}

%% file: intro_new.tex
\section{Introduction}
\label{sec:intro}

Joining tables requires identifying which values or records across sources refer to the same underlying information~\cite{halevy2006data,doan2012principles}. These correspondences support applications such as product-catalog alignment, healthcare-directory reconciliation, vendor matching, and government-record integration~\cite{kopcke2012tailoring,carvalho2015entity,malhotra2014incremental,binette2022almost}. The challenge is that these correspondences can arise in fundamentally different ways. Some columns align after simple normalization, while others require syntactic transformation, semantic matching, or record-level evidence that two rows refer to the same real-world entity~\cite{zhu2017auto,dong2021efficient,li2020deep}. For a previously unseen pair, the appropriate form of alignment is not known in advance, and the ground-truth join is unavailable when this decision must be made. Applying an unsuitable method may still produce plausible correspondences, while applying every available method incurs unnecessary computation. We refer to the problem of determining how to establish a join under such heterogeneous matching requirements as \emph{adaptive join selection}.

\begin{figure*}[t]
\centering
\includegraphics[width=\textwidth]{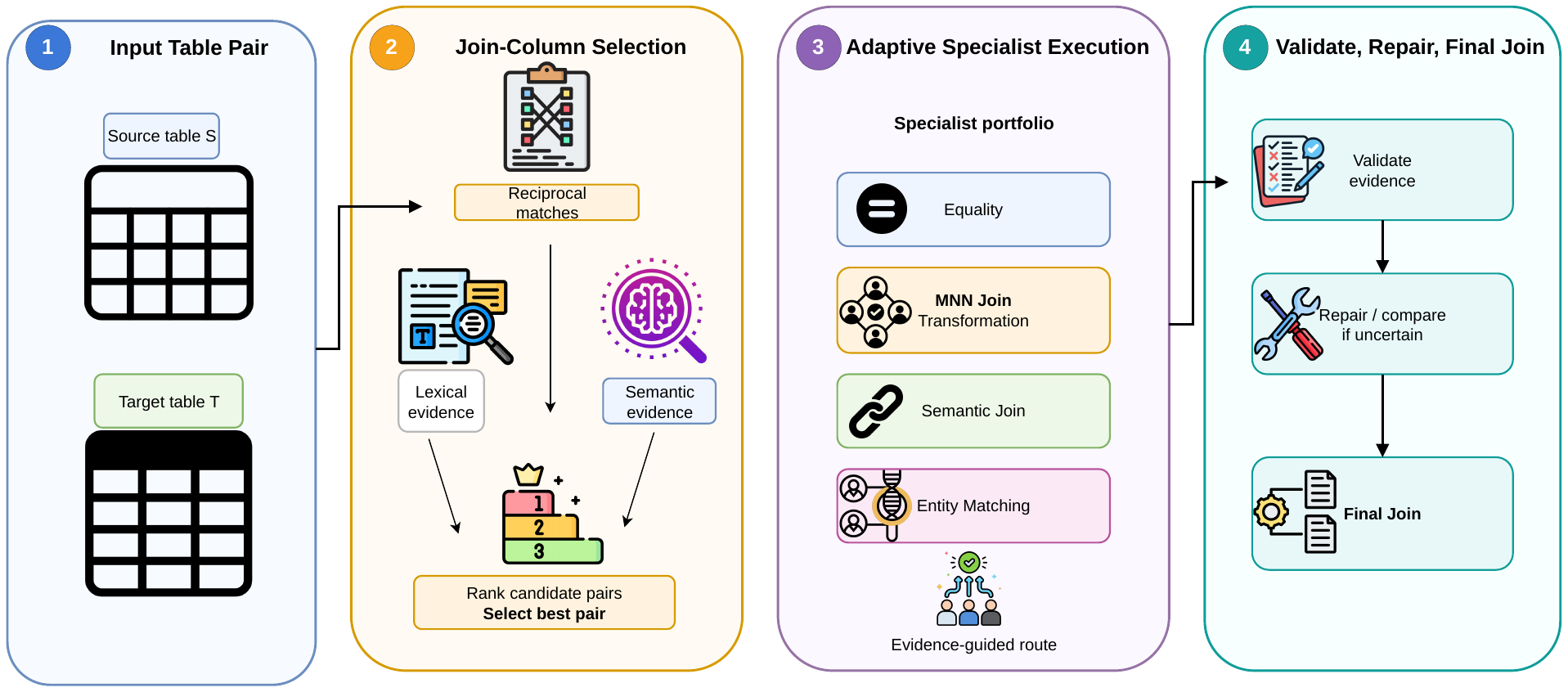}

\caption{The \textsc{MetaJoin} workflow: (1) input table pair, (2) join-column selection, (3) evidence-guided routing over the specialist portfolio, and (4) validation and repair before committing the final join.}
\label{fig:workflow}
\end{figure*}

Existing join methods address different forms of cross-source heterogeneity. Equality-based joins are inexpensive when values already share a common representation~\cite{shapiro1986join}. Transformation methods reconcile systematic differences in formatting, extraction, token order, or concatenation before equality can be applied~\cite{gulwani2011automating,singh2016blinkfill,abedjan2016dataxformer,zhu2017auto,jin2017foofah,dargahi2024dtt}. Approximate and embedding-based methods handle lexical or semantic variation~\cite{dong2021efficient,dong2022deepjoin,cong2022warpgate}, while entity-matching methods combine evidence across attributes when correspondence depends on record identity~\cite{konda2016magellan,mudgal2018deep,li2020deep,zhang2024jellyfish}. These methods differ in both the heterogeneity they can resolve and the cost of applying them, and no single method performs reliably across all settings. Running every method avoids committing to one method in advance, but repeatedly pays for transformation search, embedding computation, model inference, and pair verification even when a simpler method is sufficient. Selecting one method from input evidence alone reduces this cost, but input evidence is not always sufficient to distinguish between plausible alternatives~\cite{rice1976algorithm,xu2008satzilla}. The correspondences produced during execution provide additional evidence for determining whether the initial choice should be trusted or revised~\cite{avnur2000eddies,markl2004robust}. Adaptive join selection must therefore incorporate evidence produced during matching to confirm or revise decisions made from input evidence alone.

\begingroup\emergencystretch=1.5em  
We address this challenge with \textbf{MetaJoin}, an evidence-guided framework for \emph{adaptive join selection} over a portfolio of specialized join methods. MetaJoin realizes adaptive join selection by augmenting an initial input-based route with post-execution validation and repair. Before specialist selection, an input-only certificate first selects the join columns. MetaJoin then constructs shared \emph{input evidence}, including column statistics, an embedding scout, and the controller's own prior execution traces, to characterize the selected column pair. The controller uses this evidence to \emph{route} the pair to a primary specialist and retains a backup when the route is ambiguous. After execution, a specialist-specific \emph{output witness} \emph{validates} whether the returned correspondences provide enough evidence to commit the join. A validated result is committed, while weak or conflicting evidence triggers \emph{repair}: comparison with the backup specialist, escalation of unresolved rows to entity matching, or rerouting of rows left unmatched by a partial join. By using output evidence to confirm or revise the initial route, MetaJoin makes join selection adaptive rather than one-shot. This design preserves selective execution while allowing MetaJoin to recover from incorrect first-pass routes, and reserves expensive model-based matching and verification for cases that cheaper evidence cannot resolve. Figure~\ref{fig:workflow} summarizes this distinction between deciding \emph{which} specialist to try and deciding \emph{whether} its result should be trusted.

Evidence-guided execution depends on each specialist exposing a useful label-free indication of whether its attempted alignment succeeded. Transformation joins make this particularly difficult. Existing approaches synthesize programs from examples or generate transformed values with a model~\cite{gulwani2011automating,singh2016blinkfill,abedjan2016dataxformer,zhu2017auto,jin2017foofah,dargahi2024dtt}, but do not provide a cheap runtime criterion for determining whether an inferred transformation actually aligns two previously unaligned columns. We therefore introduce \textbf{MNN Join}, a non-LLM transformation method based on mutual-nearest-neighbor alignment. Under a useful transformation, transformed source values and their corresponding target values tend to become reciprocal nearest neighbors. Incorrect transformations yield weaker reciprocal coverage, while over-transformations collapse distinct values onto the same targets and lose reciprocity. MNN Join uses this structure to search for transformations without requiring a model to generate target values. Importantly, the same mutual-nearest-neighbor structure also provides a runtime measure of whether the resulting alignment is well supported. Mutual-nearest-neighbor coverage therefore serves as the search signal for discovering transformations, as the rule that sets the acceptance cut and emits the join, and as the output witness MetaJoin uses to validate their results.
\par\endgroup

We evaluate MetaJoin on $670$ datasets from six benchmarks, spanning a heterogeneous mix of join behaviors both across and within benchmarks. No standalone specialist performs consistently well across this workload, while \textsc{MetaJoin} achieves $0.817$ macro row-$F_1$ without benchmark or join-type labels. It slightly exceeds the $0.812$ benchmark-level oracle, which selects one best specialist per benchmark, by adapting its selection to individual joins within each benchmark. It also outperforms the strongest single join method outside its own portfolio, the per-pair LLM matcher \textsc{Lotus}~\cite{patel2024semantic}, by $0.16$ macro row-$F_1$ ($0.817$ versus $0.65$). In the transformation-focused evaluation, \textsc{MNN Join} outperforms every evaluated transformation baseline. Finally, MetaJoin varies by only $0.015$ macro row-$F_1$ across controller backbones from $3$B to $14$B, showing that validation and repair reduce dependence on the initial selection.

%% file: related.tex
\section{Related Work}
\label{sec:related}

\paragraph{Transformation-based joins and data transformation.}
Programming-by-example systems infer data transformations from demonstrated input--output pairs. \textsc{FlashFill}~\cite{gulwani2011automating} synthesizes string programs from examples, while \textsc{BlinkFill}~\cite{singh2016blinkfill} uses additional input structure to reduce example ambiguity. \textsc{Foofah}~\cite{jin2017foofah} searches over compositions of data-transformation operators, and \textsc{DataXFormer}~\cite{abedjan2016dataxformer} discovers transformations from external data. Transformation joins extend this line of work to table alignment~\cite{zhu2017auto,dargahi2024dtt,omidvartehrani2024gxjoin}. \textsc{Auto-Join}~\cite{zhu2017auto} discovers candidate correspondences between columns before synthesizing a transformation, while \textsc{DTT}~\cite{dargahi2024dtt} uses LLMs for example-driven tabular transformation. These methods infer transformations from explicit or bootstrapped correspondences. \textsc{MNN Join} instead evaluates candidate transformations directly against the two unaligned columns using reciprocal coverage, avoiding materialized input--output examples and allowing transformation search to operate as a label-free specialist within \textsc{MetaJoin}.

\paragraph{Join discovery and joinable-table search.}
Data discovery systems identify tables or columns that are likely to join. Set-based approaches use overlap or containment, including \textsc{LSH Ensemble}~\cite{zhu2016lsh} and \textsc{JOSIE}~\cite{zhu2019josie}. \textsc{Data Civilizer}~\cite{deng2017data} and \textsc{Aurum}~\cite{fernandez2018aurum} organize relationships among datasets to support discovery and join-path search. More recent methods extend joinability beyond exact overlap~\cite{li2021auto}. \textsc{PEXESO}~\cite{dong2021efficient} searches for joinable tables using high-dimensional similarity, \textsc{DeepJoin}~\cite{dong2022deepjoin} learns representations for joinable-table discovery, and \textsc{WarpGate}~\cite{cong2022warpgate} targets semantic join discovery in cloud data warehouses. This work primarily asks which columns or tables are joinable under a particular similarity model. \textsc{MetaJoin} addresses the complementary problem of how a selected column pair should be matched, choosing among heterogeneous join methods and revising that choice when execution produces contradictory evidence.

\paragraph{Semantic matching and entity resolution.}
Entity matching determines whether records refer to the same real-world object. \textsc{Magellan}~\cite{konda2016magellan} provides an entity-matching management framework, while \textsc{DeepMatcher}~\cite{mudgal2018deep} studies neural architectures for matching structured records and \textsc{Ditto}~\cite{li2020deep} applies pretrained language models to the task. Other work reduces supervision requirements. \textsc{ZeroER}~\cite{wu2020zeroer} performs entity resolution without labeled examples, \textsc{Sudowoodo}~\cite{wang2023sudowoodo} learns self-supervised representations for multiple data-integration tasks, and \textsc{Jellyfish}~\cite{zhang2024jellyfish} instruction-tunes local LLMs for data preprocessing, including entity matching. These methods improve individual matching mechanisms. \textsc{MetaJoin} treats such mechanisms as specialists and instead determines when semantic or entity-level evidence is necessary relative to cheaper equality and transformation-based alternatives.

\paragraph{Adaptive query processing and algorithm selection.}
Per-instance selection among alternative algorithms has long been studied in algorithm selection~\cite{rice1976algorithm,xu2008satzilla}, while database systems have explored both learned and runtime-adaptive optimization. \textsc{LEO}~\cite{stillger2001leo}, \textsc{Neo}~\cite{marcus2019neo}, and \textsc{Bao}~\cite{marcus2021bao} use workload or execution feedback to improve subsequent plan choices. \textsc{Eddies}~\cite{avnur2000eddies} continuously adapts execution order, progressive optimization~\cite{markl2004robust} revises plans when runtime observations disagree with estimates, and \textsc{SkinnerDB}~\cite{trummer2019skinnerdb} learns join-order decisions from execution progress. Reinforcement learning has also been used to optimize join ordering~\cite{krishnan2018learning}. \textsc{MetaJoin} shares the principle that an initial choice need not be final, but applies it to a different decision space. It selects among heterogeneous matching paradigms rather than physical plans or join orders, and uses specialist-specific output evidence to decide whether the initial selection should be retained or revised.

\paragraph{Cost-aware routing and semantic query optimization.}
Recent LLM routing systems balance model quality and inference cost. \textsc{FrugalGPT}~\cite{chen2023frugalgpt} constructs cost-aware cascades, \textsc{AutoMix}~\cite{aggarwal2024automix} combines model selection with self-verification, \textsc{HybridLLM}~\cite{ding2024hybrid} routes queries under quality and cost considerations, and \textsc{RouteLLM}~\cite{ong2025routellm} learns routing policies from preference data. \textsc{RouterBench}~\cite{hu2024routerbench} provides a benchmark for multi-LLM routing. Related data systems optimize LLM-powered operators and pipelines. \textsc{LOTUS}~\cite{patel2024semantic} exposes declarative semantic operators, \textsc{Palimpzest}~\cite{liu2025palimpzest} optimizes AI-powered analytics through declarative query processing, \textsc{Abacus}~\cite{russo2025abacus} develops cost-based optimization for semantic-operator systems, and \textsc{DocETL}~\cite{shankar2024docetl} uses agentic rewriting and evaluation for document-processing pipelines. These systems generally choose among models or implementations of a common semantic operation. \textsc{MetaJoin} instead chooses among different notions of correspondence and validates each choice using evidence specific to the selected join method.

\paragraph{LLM tool use for data systems.}
Tool-using LLMs such as \textsc{ReAct}~\cite{yao2022react} and \textsc{Toolformer}~\cite{schick2023toolformer} show how language models can interleave reasoning with external actions, while recent work such as \textsc{TAG}~\cite{biswal2024text2sql} explores tighter integration between language models and database execution. \textsc{MetaJoin} uses an LLM in a narrower orchestration role. Input evidence establishes an initial specialist selection, execution produces method-specific output evidence, and validation or repair determines whether that selection should be retained. The final join therefore depends on evidence from the matching process rather than an unchecked model decision.

%% file: method.tex
\section{Method}
\label{sec:method}

\subsection{Problem Formulation}
\label{sec:problem}

A \emph{table-join task} consists of a source table $\mathcal{S}$ and a target
table $\mathcal{T}$. The policy must choose a source column $a\in\mathcal{S}$,
a target column $b\in\mathcal{T}$, and a join over their values. For a chosen
pair $(a,b)$, let $S=\{s_1,\dots,s_m\}$ and $T=\{t_1,\dots,t_n\}$ be the
corresponding multisets of strings. The ground-truth relation
$J^\star_{a,b}\subseteq S\times T$ contains the source-target value pairs that
denote the same real-world quantity, and is empty when $(a,b)$ is not joinable.

Joinable pairs differ in the evidence that supports their matches. Some are
exact after trivial normalization, some require a regular string transformation,
some require semantic equivalence between different surface forms, and some
require entity-level evidence from record descriptions. We use these four
behaviors only to describe the workload and specialist portfolio. They are not
observed by the policy.

A \emph{specialist} is a procedure $M$ that takes a selected column pair $(S,T)$
and returns a predicted join $\hat J_M=M(S,T)\subseteq S\times T$ at cost
$c(M,S,T)$. We measure cost by wall-clock time and model tokens,
$c=(c_{\mathrm{time}},c_{\mathrm{tok}})$. A specialist portfolio
$\mathcal{M}=\{M_1,\dots,M_K\}$ may include cheap symbolic specialists, such as
equality or transformation search, and more expensive model-based specialists,
such as semantic or entity matching.

Given a table pair $(\mathcal{S},\mathcal{T})$ and a specialist portfolio
$\mathcal{M}$, the adaptive join-selection problem is to learn a policy $\pi$ that
selects a candidate column pair, invokes one or more specialists, and commits
a final join $\hat{J}_{\pi}$. \textsc{MetaJoin} targets the Pareto frontier
between expected join quality and execution cost:
\begin{equation}
\label{eq:objective}
\max_{\pi}\;
\mathbb{E}_{(\mathcal{S},\mathcal{T})}
\!\left[
F_1\!\left(\hat{J}_{\pi},J^\star\right)
\right],
\qquad
\min_{\pi}\;
\mathbb{E}_{(\mathcal{S},\mathcal{T})}
\!\left[
c_{\pi}(\mathcal{S},\mathcal{T})
\right].
\end{equation}
Here, $J^\star$ denotes the latent target join, and
$c_{\pi}=(c_{\mathrm{time}},c_{\mathrm{tok}})$ captures wall-clock and model
cost. \textsc{MetaJoin} approaches this objective through selective execution,
invoking additional specialists or verification only when the current evidence
is insufficient.

\subsection{Selecting the Join Columns}
\label{sec:colpick}

Since every specialist operates on the selected columns, this choice bounds the
quality of the final join. A wrong column pair may contain no recoverable
correspondence, while a non-optimal pair may preserve only partial or noisy
evidence and thus limit attainable recall or precision. The selector must
therefore identify joinability rather than simply choose the pair with the
largest raw string overlap. Single overlap scores are fragile: repeated tokens
create hubs, semantically related columns need not be joinable, and true
transform joins may have weak overlap before the transformation is found.
\textsc{MetaJoin} therefore scores each candidate pair $(a,b)$ by an
input-only certificate built from hub-free reciprocal value matches.

For candidate columns $a$ and $b$, let
$\Sigma\in\mathbb{R}^{n_a\times n_b}$ be the value-similarity matrix, where $\Sigma_{ij}$ is a normalized string-overlap score between value
$u_i\in a$ and value $v_j\in b$.  A value pair $(u_i,v_j)$ is a reciprocal match if
each value is the other's strict nearest neighbor:
\[
\Sigma_{ij}>\max_{j'\neq j}\Sigma_{ij'}
\quad\wedge\quad
\Sigma_{ij}>\max_{i'\neq i}\Sigma_{i'j}.
\]

Strict inequalities exclude ties. Let $M(a,b)$ be the reciprocal-match set and
$m(a,b)=|M(a,b)|$. Reciprocity prevents repeated or generic values from
inflating the evidence, since many rows may point to the same candidate but only one can be reciprocated.

The certificate has two witnesses. For $(i,j)\in M(a,b)$, let
$x_{ij}=\mathrm{anchor}(i,j)$ be the shared token span responsible for the
string-overlap score, and let $\mathrm{freq}_a(x)$ and $\mathrm{freq}_b(x)$
denote the number of values in columns $a$ and $b$ that contain span $x$. The
lexical witness counts reciprocal matches whose anchor is unique on both sides:
\[
\mathrm{ev}(a,b)=
\left|
\left\{
(i,j)\in M(a,b)
\mid
\mathrm{freq}_a(x_{ij})=\mathrm{freq}_b(x_{ij})=1
\right\}
\right|.
\]

The semantic witness rechecks the same anchors in embedding space. Let
$\mathrm{sem}(i,j)=1$ iff the anchors of $u_i$ and $v_j$ are strict reciprocal
nearest neighbors after embedding. Then
\[
\ell(a,b)=
\left|
\left\{
(i,j)\in M(a,b)
\mid
\mathrm{sem}(i,j)=1
\right\}
\right|.
\]
Thus, $\mathrm{ev}$ removes evidence from repeated spans, while $\ell$ removes
evidence from anchors that do not remain coherent beyond string overlap.

Each column pair receives a tier and an evidence count:
\[
(\mathrm{tier},e)(a,b)=
\begin{cases}
(\textsc{hybrid},\min(\mathrm{ev},\ell)) &
\mathrm{ev}>0,\ \ell>0,\\
(\textsc{lexical},\mathrm{ev}) &
\mathrm{ev}>0,\ \ell=0,\\
(\textsc{sem-rescue},\ell) &
\mathrm{ev}=0,\ \ell>0,\ m(a,b)<r,\\
(\textsc{none},0) &
\text{otherwise},
\end{cases}
\]
where $r=3$ marks the sparse-reciprocity regime. We restrict
\textsc{sem-rescue} to this regime because semantic anchors are most useful when lexical evidence is absent and only a few reciprocal matches are available.

The selector ranks candidate pairs lexicographically by tier,
\[
\textsc{hybrid}
\succ
\textsc{lexical}
\succ
\textsc{sem-rescue}
\succ
\textsc{none},
\]
then by the evidence score $e(a,b)$, and finally by the number of reciprocal
matches $m(a,b)$. The highest-ranked column pair is selected once and passed
unchanged to all downstream specialists.

\subsection{Specialists and Runtime Evidence}
\label{sec:specialists}

The selected columns may require different forms of evidence to join. Direct
overlap is best handled by cheap symbolic matching, regular rewrites require
program search, semantic equivalence requires semantic matching, and descriptive
entity values require entity matching. A single matcher either overpays on easy
cases or becomes too permissive on hard ones. \textsc{MetaJoin} therefore uses a
portfolio of specialists and lets the controller choose which one to invoke. A specialist receives the selected column pair $(S,T)$ and returns a candidate
join $P_M\subseteq S\times T$, together with its cost and any evidence needed for
verification. The portfolio contains four specialist classes:
normalization-based equality matching, transformation search, semantic matching,
and entity matching. The transformation specialist is MNN Join
(\S\ref{sec:mnnjoin}), which searches for an explicit string program, then
commits the reciprocal matches its transform induces; the semantic specialist is \textsc{Semantic Join} (\S\ref{sec:semjoin}), a label-free embedding nearest-neighbor matcher. These classes are actions available to the controller.

Before invoking a specialist, the controller summarizes the selected columns
with a shared, label-free evidence profile,
\[
\phi(S,T)=
\bigl(
\phi_{\mathrm{overlap}},
\phi_{\mathrm{format}},
\phi_{\mathrm{structure}},
\phi_{\mathrm{size}},
\phi_{\mathrm{sem}}
\bigr).
\]
The first four groups capture direct string overlap, value-format regularity,
duplicate and multiword structure, and column cardinality. These signals provide
complementary views of the candidate join and are interpreted jointly by the
routing policy against the available specialist catalog
(\S\ref{sec:policy}).

The semantic component is computed by an embedding scout. For each
source value $s\in S$, let
\[
\sigma_1(s)
=
\max_{t\in T}
\cos\!\bigl(e(s),e(t)\bigr),
\]
where $e(\cdot)$ is a sentence encoder, and let $\sigma_2(s)$ denote the
second-highest target similarity. The scout records quantiles of the top-1
scores $\sigma_1(s)$ together with the average nearest-neighbor margin,
\[
\operatorname{mgn}(S,T)
=
\frac{1}{|S|}
\sum_{s\in S}
\bigl(\sigma_1(s)-\sigma_2(s)\bigr).
\]
High top-1 similarities with clear margins indicate a coherent alignment,
whereas weak or flat score distributions suggest ambiguity. The controller
uses this information together with the remaining input evidence to select a
primary specialist and a backup when necessary.

After a specialist $M$ returns candidate pairs
$P_M\subseteq S\times T$, the controller augments the shared state with
runtime output evidence $q_M$, the specialist's \emph{output witness}. This evidence is derived from the returned
matches and may include exact agreement, reciprocal structure, similarity-score
separation, sampled pair judgments, or agreement with another specialist. The
controller interprets these signals jointly when deciding whether to commit the
result, compare alternatives, or escalate.

It also records two operator-independent diagnostics:

\[
  {\operatorname{srcCov}(P)=\frac{|\{s:\exists t,\,(s,t)\in P\}|}{|S|},
  \quad
  \operatorname{bij}(P)=\frac{|\{t:\exists s,\,(s,t)\in P\}|}{|P|}.
}\]
Source coverage measures how much of the source column is explained by the
candidate join, while bijectivity summarizes the extent of many-to-one
collisions. These quantities describe the size and structure of the result, but
do not serve as standalone acceptance criteria. Input evidence therefore guides
which specialist to invoke, while the augmented runtime evidence determines
whether to commit, compare, or escalate its output.

\subsection{MNN Join as the Transformation Specialist}
\label{sec:mnnjoin}

When the selected columns are routed to the transformation specialist, the system needs one that is accurate enough to recover representation changes, cheap enough to run inside a cost-aware portfolio, label-free, and able to report at runtime whether the transformation it found actually aligns the two columns. This rules out two natural extremes. Scoring candidate programs only by exact matches is cheap, but too sparse to guide search because useful partial transformations may create no exact equality until the final operator is found. Model-based generation can propose transformations more flexibly, but repeated calls make it expensive as a specialist that may be invoked across many table pairs. MNN Join performs a non-LLM search over string programs, using a label-free alignment score that is dense enough to guide partial programs but structured enough to avoid many-to-one collapse.

Let $\Omega$ be a library of unary transformation operators, including substring extraction, delimiter splitting, token selection and reordering, duplicate removal, sorting, prefix and initial extraction, parenthetical or numeric stripping, and concatenation with auxiliary columns (\Cref{tab:operators} lists the full $25$-operator library). A transformation chain is a
composition of these operators,
\[
  \tau=\omega_k\circ\cdots\circ\omega_1,
  \qquad \omega_i\in\Omega .
\]
Operators marked \emph{repeatable} may be applied more than once within a single search step (e.g., peeling several leading characters), which lets a one-step extension express a bounded family of fixed-length edits without enlarging the branching factor.

\begin{table*}[t]
\centering
\caption{The MNN Join operator library $\Omega$: $25$ unary string operators, listed
two per row. $^\dagger$\,repeatable within one search step.}
\label{tab:operators}
\small
\setlength{\tabcolsep}{5pt}
\begin{tabular}{@{}ll@{\hspace{1.4em}}ll@{}}
\toprule
Operator & Example & Operator & Example \\
\midrule
\multicolumn{4}{@{}l}{\textit{Substring / character extraction (char)}}\\
auto-split, keep field   & \texttt{a:b$\to$b}      & drop leading char(s)$^\dagger$  & \texttt{Xenon$\to$enon} \\
drop trailing char(s)$^\dagger$ & \texttt{Smiths$\to$Smith} & drop first char        & \texttt{\_id$\to$id} \\
drop last char           & \texttt{id\_$\to$id}    & remove 2nd char         & \texttt{u.s$\to$us} \\
remove 2nd-to-last char  & \texttt{abXc$\to$abc}   & take leading prefix     & \texttt{Robert$\to$Rob} \\
strip parenthetical      & \texttt{Ann\,(NE)$\to$Ann} & strip leading numeric & \texttt{12.\,Ann$\to$Ann} \\
\midrule
\multicolumn{4}{@{}l}{\textit{Delimiter splitting + field selection (token)}}\\
split `-', keep first    & \texttt{a-b-c$\to$a}    & split `-', keep last    & \texttt{a-b-c$\to$c} \\
split space, keep first  & \texttt{J Smith$\to$J}  & split space, keep last  & \texttt{J Smith$\to$Smith} \\
keep part after `-'      & \texttt{x-42$\to$42}    & keep $k$-th token, front$^\dagger$ & \texttt{a b c$\to$a} \\
keep $k$-th token, back$^\dagger$ & \texttt{a b c$\to$c} &                    &  \\
\midrule
\multicolumn{4}{@{}l}{\textit{Token reordering / normalization (token)}}\\
cyclic word shift        & \texttt{a b c$\to$c a b} & move first token last   & \texttt{J Smith$\to$Smith J} \\
sort tokens              & \texttt{b a$\to$a b}    & remove duplicate tokens & \texttt{ny ny$\to$ny} \\
keep alphabetic tokens   & \texttt{Apt 4B$\to$Apt} & initials joined by `.'   & \texttt{John Smith$\to$J.S.} \\
\midrule
\multicolumn{4}{@{}l}{\textit{Concatenation with an auxiliary column (concat)}}\\
prepend another column   & \texttt{Smith$\to$John Smith} & append another column & \texttt{John$\to$John Smith} \\
\bottomrule
\end{tabular}
\end{table*}

MNN Join replaces exact-match feedback during search with a label-free
reciprocal-alignment score. Each value $x$ is represented by an IDF-weighted
character $n$-gram vector $\psi(x)$, with IDF estimated from the selected source
and target columns. Let
\[
  \kappa(x,y)=\cos(\psi(x),\psi(y)).
\]
For a chain $\tau$, define the source-target similarity matrix
\[
  K_{ij}=\kappa\bigl(\nu(\tau(s_i)),\nu(t_j)\bigr).
\]
A pair $(s_i,t_j)$ is mutual nearest under $K$ when $t_j$ is the most similar
target for $s_i$ and $s_i$ is the most similar source for $t_j$. MNN Join scores
$\tau$ by the fraction of source values that participate in such reciprocal
matches (the mutual-nearest-neighbor coverage):
\[
  A(\tau)=
  \frac{1}{|S|}
  \left|
  \left\{
  i:\exists j\ \mathrm{s.t.}\ (s_i,t_j)\ \text{is mutual nearest under }K
  \right\}
  \right|.
\]

The score favors transformations that make corresponding values both similar and
one-to-one. This distinction matters because average similarity alone can reward
over-transformation. If many source values collapse to the same target
representation, only one can remain reciprocal, so $A(\tau)$ drops without a
separate collision penalty. MNN Join grows the chain greedily from the identity transformation. At each step,
it evaluates every one-operator extension of the current chain and applies the
operator that gives the largest strict increase in $A(\tau)$. The search stops
when no extension improves the score. The procedure is run in both directions,
$S\to T$ and $T\to S$, and the higher-scoring direction is retained.

After the search fixes $\tau^\star$, MNN Join forms the join from the reciprocal
alignment that $\tau^\star$ induces, not from exact string equality. Let
$K^\star_{ij}=\kappa\bigl(\nu(\tau^\star(s_i)),\nu(t_j)\bigr)$ be the final
similarity matrix. 

For each source value $s_i$, let
\[
  r_i=\max_j K^\star_{ij},
  \qquad
  j(i)=\arg\max_j K^\star_{ij}.
\]
Let $\mathcal{R}$ be the set of source rows whose best match is reciprocal,
\[
  \mathcal{R}
  =
  \{\,i:(s_i,t_{j(i)})\ \text{is mutual nearest under }K^\star\,\}.
\]
For a threshold $\theta$, let
\[
  \mathcal{C}_\theta=\{\,i:r_i\ge \theta\,\}
\]
be the source rows admitted by best-match similarity. MNN Join chooses the
threshold by comparing this admitted set with the reciprocal set:
\[
  \rho(\theta)
  =
  \frac{2|\mathcal{C}_\theta\cap\mathcal{R}|}{|\mathcal{C}_\theta|+|\mathcal{R}|}.
\]
Although $\rho$ has the algebraic form of an $F_1$ score, both sets are derived
only from $K^\star$. It is therefore a label-free reciprocity agreement, not an
evaluation against ground truth. MNN Join sets $\theta^\star$ to the lowest
observed value of $r_i$ that maximizes $\rho(\theta)$, subject to a small minimum
floor.

The committed join is the thresholded reciprocal alignment under $\tau^\star$:
\[
  \hat J_{\tau^\star}
  =
  \{\, (s_i,t_{j(i)}) :
      i\in\mathcal{R},\ r_i\ge\theta^\star \,\}.
\]
This commitment rule is less brittle than requiring transformed equality
$\nu(\tau^\star(s))=\nu(t)$. A correct transformation may still leave residual
punctuation, abbreviation, or token-order differences that prevent exact equality.
Thresholding keeps such near-matches, while reciprocity preserves one-to-one
structure. Thus the same mutual-nearest principle selects the chain, sets the
cut, and emits the final pairs without labels.

\subsection{Semantic Join: the Semantic Specialist}
\label{sec:semjoin}

\textsc{Semantic Join} covers two cases that transformation search may not resolve: column pairs related primarily by meaning, and residual transformation cases whose surface rewrites are too irregular, or outside the operator library, but still preserve semantic similarity. It embeds each source and target value with a general-purpose sentence encoder, $L_2$-normalizes the embeddings, and matches each source value to its nearest target by cosine similarity. The controller filters forced matches in sparse or ambiguous joins using the observed score distribution (\S\ref{sec:policy}).

\subsection{Routing, Validation, and Repair}
\label{sec:policy}

Join tasks are heterogeneous, and their evidence profiles often overlap across
specialists. A fixed policy may therefore select a poorly suited specialist,
while evaluating many specialists incurs unnecessary cost. \textsc{MetaJoin}
addresses this tradeoff through evidence-guided execution. For each selected
column pair, it first \emph{routes} using shared input evidence, then
\emph{validates} the returned candidate join using runtime output evidence, and
finally \emph{repairs} an inconclusive result through comparison, escalation,
or residual rerouting. Output evidence thus decides whether the initial route is trusted or revised.

For a selected pair $(S,T)$, the initial controller state is
\[
h_0 =
\bigl(
\phi(S,T),
\eta(S,T),
r(S,T)
\bigr),
\]
where $\phi(S,T)$ is the shared input profile, $\eta(S,T)$ is the
embedding-scout summary, and $r(S,T)$ is the experience prior. Each specialist
invocation augments this state with
\[
o_t =
\bigl(
M_t,
P_t,
\operatorname{srcCov}(P_t),
\operatorname{bij}(P_t),
q_t
\bigr),
\qquad
h_{t+1}=h_t\cup\{o_t\},
\]
where $P_t$ is the returned candidate join and $q_t$ denotes the runtime
evidence derived from that result. Algorithm~\ref{alg:controller} summarizes
the resulting execution.

\begin{algorithm}[t]
\small
\caption{Evidence-guided execution for a selected column pair $(S,T)$.}
\label{alg:controller}
\begin{algorithmic}
\Require selected pair $(S,T)$, input profile $\phi$, scout summary $\eta$,
         experience prior $r$, specialist portfolio $\mathcal{M}$
\State $(M_1,M_2)\gets{}$ \Call{Route}{$\phi,\eta,r,\mathcal{M}$}
\State $P_1\gets M_1(S,T)$
\State $q_1\gets{}$ \Call{Evaluate}{$M_1,P_1$}
\State $\hat J\gets{}$ \Call{Commit}{$P_1,q_1$}
\If{$M_2\neq\varnothing$}
    \State $P_2\gets M_2(S,T)$
    \State $q_2\gets{}$ \Call{Evaluate}{$M_2,P_2$}
    \State $\hat J\gets{}$ \Call{CompareAndFuse}{$M_1,P_1,q_1,M_2,P_2,q_2$}
\EndIf
\State $\hat J\gets{}$ \Call{EntityRepair}{$S,T,\hat J,\mathcal{M}$}
\State $S_{\mathrm{res}}\gets S\setminus\{s:\exists t,\,(s,t)\in\hat J\}$
\While{$S_{\mathrm{res}}\neq\varnothing$ and an unused specialist remains}
    \State $\hat J\gets \hat J\cup{}$ \Call{Reroute}{$S_{\mathrm{res}},T,\mathcal{M}$}
    \State update $S_{\mathrm{res}}$
\EndWhile
\State append the witness-supported execution trace to $r$
\State \Return $\hat J$
\end{algorithmic}
\end{algorithm}

\paragraph{Route.}
Routing chooses the first specialist to try from label-free evidence. The catalog reader receives the input evidence $\phi(S,T)$, the embedding-scout summary $\eta(S,T)$, the experience prior $r$, and the specialist portfolio $\mathcal{M}$:
\[
(M_1,M_2)=R_{\mathrm{read}}\!\bigl(
\phi(S,T),\eta(S,T),r(S,T),\mathcal{M} \bigr),
\]
where $M_1$ is the primary route and $M_2$ is an optional backup when multiple specialists remain plausible. The reader interprets the measured evidence jointly with the specialist descriptions. Near-identical values favor equality matching. Regular character-level differences suggest transformation search, while lexical variation with semantic agreement points to semantic matching. Structured multiword records instead favor entity matching.

\paragraph{Experience prior.} Join workloads are repetitive: similar column pairs recur across tables, so routing each from scratch repeats deliberation the controller has already paid for. The prior $r(S,T)$ amortizes that by reusing the controller's own past evidence. After a task finishes, \textsc{MetaJoin} stores its input evidence, invoked specialists, output evidence, accepted action, and escalation decision. For a new task, it retrieves nearby past tasks in evidence space and estimates \[ r_M(S,T)= \frac{1}{|\mathcal{N}_k(S,T)|} \sum_{j\in\mathcal{N}_k(S,T)} q_M^{(j)}, \] where $q_M^{(j)}$ is the previous label-free witness for specialist $M$.

\paragraph{Validate.}
Validation augments the controller state with runtime evidence from the returned
candidate join. This evidence includes the specialist's own output signals,
generic diagnostics such as coverage and bijectivity, and, when needed, sampled
pair judgments from the LLM pair verifier or agreement with another specialist. Different specialists
contribute different forms of evidence. Equality may provide normalized exact
agreement, while \textsc{MNN Join} exposes the reciprocal alignment induced by
its selected transformation. Semantic and entity candidates additionally
provide similarity structure and record-level evidence. These signals determine whether the candidate is committed directly or passed
to repair.

\paragraph{Repair: arbitration.}
When both the primary and backup specialists are executed, \textsc{MetaJoin} reconciles their candidate joins before committing. Let \(P_1\) and \(P_2\) denote their outputs, with \(P_{\cap}=P_1\cap P_2\) containing pairs proposed by both specialists and \(P_{\Delta}=(P_1\cup P_2)\setminus P_{\cap}\) containing pairs proposed by only one. Agreed pairs receive stronger support, while disputed pairs are evaluated using the same class-specific witnesses. This comparison is particularly important for specialists prone to over-merging, such as entity matching, where multiple source values may be assigned to the same target. If two pairs \((s,t)\) and \((s',t)\) compete for the same target, the controller retains the pair supported by both specialists and rejects an alternative proposed by only one. Cross-specialist agreement therefore filters likely false positives. The final join is \(P_{\mathrm{fuse}}=P_{\cap}^{\mathrm{acc}}\cup\mathrm{Accept}(P_{\Delta})\), where \(P_{\cap}^{\mathrm{acc}}\) contains agreed pairs that also pass the relevant evidence checks.

\paragraph{Repair: escalation and residuals.}
The initial route may be poorly suited to the observed column pair. \textsc{MetaJoin}
therefore compares joins generated by each candidate using the runtime evidence accumulated for
each result rather than a single universal score. For transformation-oriented
routes, program-induced reciprocal alignment provides the main comparative
signal. For meaning-based routes, verified agreements are retained, while
contested pairs are passed to arbitration. A gap between raw and reciprocal coverage indicates that apparent surface
matches do not form a consistent transformation. The controller then escalates
the unresolved rows to entity matching, where record-level evidence can
disambiguate them.

The policy combines relative comparisons, cross-specialist agreement, and cutoffs derived from the current table. Finally, source rows not covered by the
committed partial join,
\[
S_{\mathrm{res}}
=
S\setminus
\{s:\exists t,\,(s,t)\in P\},
\]
are routed again through the remaining specialists. This allows different
subsets of the same table pair to be handled by different operators.

%% file: exp_new.tex
\section{Experiments}
\label{sec:experiments}

\paragraph{Overview.}
Our evaluation addresses three questions. \textbf{RQ1} asks whether adaptive join selection improves end-to-end accuracy over fixed specialists, benchmark-level selection, and label-free one-shot alternatives on a heterogeneous workload (\S\ref{sec:end2end}). \textbf{RQ2} identifies the signals and mechanisms that make this selection adaptive, including column selection, transformation search, input-side routing, output validation and repair, experience reuse, and portfolio composition (\S\ref{sec:mechanisms}). \textbf{RQ3} examines whether adaptive selection retains the efficiency benefits of selective execution and remains scalable and robust as table size and controller backbone change (\S\ref{sec:deployment}). All methods follow the same label-free protocol (\S\ref{sec:setup}).

\subsection{Setup and Label-Free Protocol}
\label{sec:setup}

\begin{figure}[t]
\centering
\includegraphics[width=\columnwidth]{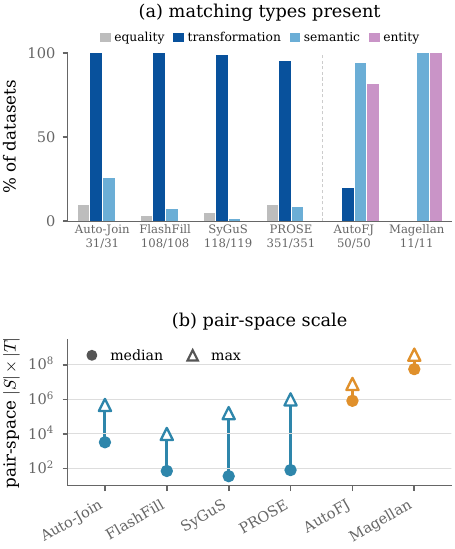}
\caption{(a) Matching behaviors present in each benchmark, manually characterized from gold pairs. (b) Median and maximum source--target pair space.}
\label{fig:benchtypes}
\end{figure}

\paragraph{Workload and matching heterogeneity.}
\begingroup\emergencystretch=1.5em  
We evaluate \textsc{MetaJoin} on $670$ datasets from six established benchmarks: \textsc{Auto-Join} ($31$)~\cite{zhu2017auto}, \textsc{FlashFill} ($108$)~\cite{gulwani2011automating}, \textsc{SyGuS} ($119$)~\cite{alur2016sygus}, \textsc{PROSE} ($351$)~\cite{microsoftprosebenchmarks}, \textsc{AutoFJ} ($50$)~\cite{li2021auto}, and \textsc{Magellan} ($11$)~\cite{konda2016magellan}. We manually characterize each dataset from ground-truth pairs using four non-exclusive classes: \emph{equality} when values match after case and whitespace normalization, \emph{transformation} when one consistent string or numeric pattern explains the pairs, \emph{semantic} when the target cannot be produced from the source characters by such a transformation, and \emph{entity} when matching requires reconciling multiple noisy attributes referring to the same real-world object. Figure~\ref{fig:benchtypes}(a) shows substantial heterogeneity both across and within benchmarks. Transformation is present in $96$--$100\%$ of the four transformation-discovery suites, yet some datasets require lookup-like semantic mappings, while semantic and entity behavior overlap on $92\%$ of \textsc{AutoFJ} and all \textsc{Magellan} tables. This heterogeneity therefore motivates adaptive selection at the individual-join level.
\par\endgroup

\paragraph{Evaluation protocol.}
\begingroup\emergencystretch=1.5em  
Unless otherwise stated, the controller and pair verifier use \textsc{Qwen3-8B-Instruct}  served with vLLM, while the entity specialist uses  \textsc{Jellyfish-8B}~\cite{zhang2024jellyfish}. All LLM-based join methods (\textsc{DTT}~\cite{dargahi2024dtt}, \textsc{Lotus}~\cite{patel2024semantic}, and \textsc{FM-Wrangle}~\cite{narayan2022can}) run on the same \textsc{Qwen3-8B-Instruct} backbone, so accuracy differences reflect the method rather than the underlying model.  Our primary metric is macro row-$F_1$, averaged equally over the six benchmarks; we also report wall-clock time, token usage, model calls, and specialist invocations. Unless explicitly labeled ``pooled,'' all aggregate results use this  benchmark macro average. Ground-truth joins are hidden during column selection, routing, threshold selection, validation, and repair, and are used only for final evaluation. Figure~\ref{fig:benchtypes}(b) summarizes workload scale, with the largest dataset containing $386$ million candidate pairs.
\par\endgroup

\subsection{RQ1: End-to-End Adaptive Join Selection}
\label{sec:end2end}

We first ask whether adaptive join selection improves end-to-end performance across the full heterogeneous workload. We compare \textsc{MetaJoin} with standalone specialists, a benchmark-level specialist oracle, and label-free policies that either select once from input evidence or run the entire portfolio before choosing an output.

\input{figures/master_table}

\paragraph{Performance across heterogeneous joins.}
Table~\ref{tab:results} shows that specialist performance varies substantially across benchmarks. On \textsc{Auto-Join}, \textsc{FlashFill}, \textsc{SyGuS}, and \textsc{PROSE}, which our workload characterization identifies as transformation-dominant, \textsc{MNN Join} is the strongest specialist and outperforms the best competing method by $0.05$--$0.15$ row-$F_1$. This advantage does not transfer to \textsc{AutoFJ} and \textsc{Magellan}, where \textsc{MNN Join} drops to $0.36$ and $0.00$, while the entity matcher reaches $0.66$ and $0.57$. \textsc{MetaJoin} instead remains close to the strongest specialist across all six benchmarks, yielding a macro row-$F_1$ of $0.82$, compared with $0.67$ for the strongest standalone specialist and $0.65$ for the strongest external baseline, \textsc{Lotus}~\cite{patel2024semantic}.

\begin{figure*}[t]
\centering
\includegraphics[width=\textwidth]{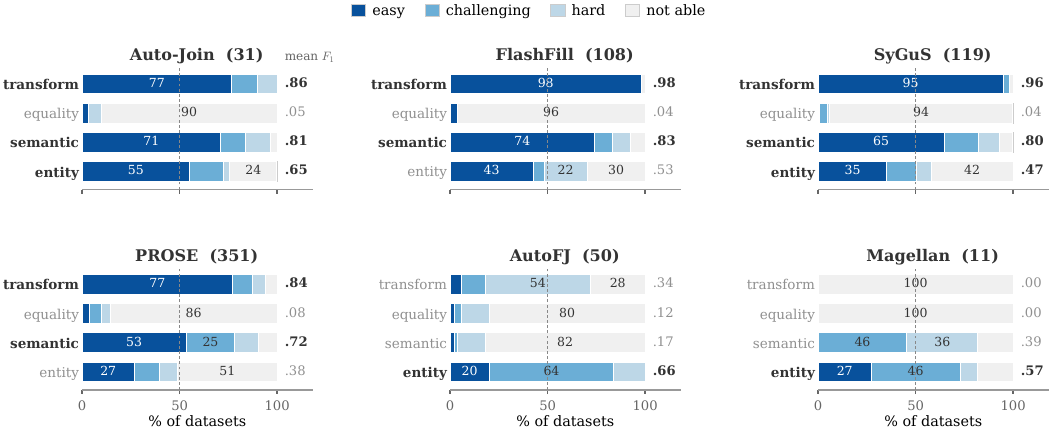}
\caption{Per-dataset capability of each matching class. For each class, we bin the row-$F_1$ of its strongest standalone specialist as \emph{easy} ($\geq.80$), \emph{challenging} ($.50$--$.80$), \emph{hard} ($.20$--$.50$), or \emph{not able} ($<.20$).}
\label{fig:difficulty}
\end{figure*}

\paragraph{Why benchmark-level selection is insufficient.}
\textsc{MetaJoin} reaches $0.817$ macro row-$F_1$, slightly exceeding the $0.812$ benchmark-level specialist oracle that selects the best single standalone specialist for each benchmark. This is possible because the best matching method can vary across individual joins within the same benchmark. Figure~\ref{fig:difficulty} makes this variation explicit. Transformation and semantic specialists both reach at least $0.50$ row-$F_1$ on $80\%$ of \textsc{Auto-Join}, $83\%$ of \textsc{FlashFill}, $82\%$ of \textsc{SyGuS}, and $74\%$ of \textsc{PROSE} datasets, while only a small fraction are served exclusively by either class. Capability also varies within a class. On \textsc{PROSE}, transformation is easy on $77\%$ of datasets but hard or worse on $13\%$. A benchmark-level policy therefore inherits the failures of its chosen specialist on joins better handled by another method. Two cases show how \textsc{MetaJoin} adapts at this finer granularity. On \textsc{Auto-Join}'s \texttt{park to state}, a transformation-based route reaches only $0.491$ row-$F_1$, while input evidence shows strong semantic agreement despite weak character overlap and routes the join to \textsc{Semantic Join}, raising row-$F_1$ to $0.909$. On \textsc{PROSE}'s \texttt{DateTime.000102}, the initial \textsc{MNN Join} route reaches only $0.059$ on a target that retains the hour of each timestamp. Its weak output witness leaves the selection unresolved, so \textsc{MetaJoin} compares the backup \textsc{DTT} output and commits its join at $0.921$. Both outcomes are identical across three seeds. These examples illustrate the two levels of adaptation in \textsc{MetaJoin}. Input evidence can select a more suitable specialist before execution, while output evidence can revise an initial selection when its result is not sufficiently supported.

Figure~\ref{fig:difficulty} resolves this effect at the dataset level. Matching classes overlap substantially rather than partitioning the workload. Transformation, semantic, and entity specialists can each solve subsets of the same benchmarks, and their capability varies within a benchmark. For example, transformation and semantic specialists both reach at least $0.50$ row-$F_1$ on $80\%$ of \textsc{Auto-Join}, $83\%$ of \textsc{FlashFill}, $82\%$ of \textsc{SyGuS}, and $74\%$ of \textsc{PROSE} datasets, while at most $16\%$ are served only by transformation and at most $4\%$ only by semantic matching. Within \textsc{PROSE}, transformation is easy on $77\%$ of datasets but hard or worse on $13\%$. The real-world suites exhibit a different overlap pattern: on \textsc{AutoFJ}, only $22\%$ of datasets admit two or more capable classes and $62\%$ are served by the entity matcher alone. A policy that fixes one specialist for an entire benchmark therefore inherits that specialist's failures on the datasets it cannot serve.

\paragraph{Why one-shot selection is insufficient.}
The two label-free policies in Table~\ref{tab:results} isolate the value of specialist-specific output evidence. The LLM router uses the same decision model as \textsc{MetaJoin}, but selects one specialist from raw column samples alone and commits its output. It reaches only $0.679$ macro row-$F_1$ and falls to $0.15$ on \textsc{AutoFJ} and $0.29$ on \textsc{Magellan}. Its $0.1$k specialist tokens per dataset are also consistent with rarely invoking the expensive entity matcher. Run-all-select removes this limitation by executing every specialist before asking the same model to choose among their outputs. Yet despite spending about $193$k specialist tokens per dataset, it reaches only $0.659$ macro row-$F_1$, including $0.24$ on \textsc{AutoFJ} and $0.44$ on \textsc{Magellan}. This suggests that increasing the choice space makes reliable selection more difficult for the model. Thus, neither input-only routing nor access to all candidate outputs is sufficient for reliable selection. \textsc{MetaJoin} exceeds both policies by more than $0.13$ by using specialist-specific output witnesses to distinguish plausible alternatives and revise the initial selection when needed.

\subsection{RQ2: Signals and Mechanisms for Adaptive Selection}
\label{sec:mechanisms}

We next isolate the signals and mechanisms that make \textsc{MetaJoin}'s selection adaptive. We first evaluate the input-only column selector and \textsc{MNN Join}'s reciprocal-coverage signal independently. We then examine how input evidence establishes the initial specialist route and how output validation and repair can revise it. Finally, we study experience reuse and sensitivity to the available specialist portfolio.

\subsubsection{Standalone Components and Signals}
\label{sec:specialist}

\begin{figure*}[t]
\centering
\includegraphics[width=\textwidth]{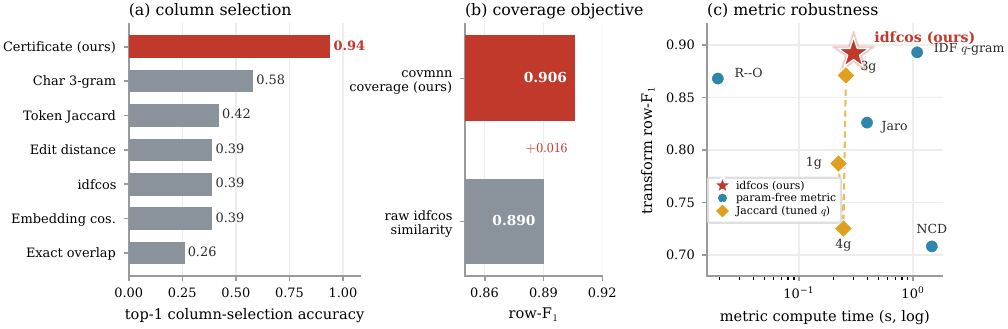}
\caption{Standalone components and signals. (a) Input-only certificate column selection versus individual signals on \textsc{Auto-Join}. (b) Reciprocal \textsc{covmnn} coverage versus raw \textsc{idfcos} similarity. (c) Sensitivity to the underlying similarity metric.}
\label{fig:standalone}
\end{figure*}

\paragraph{Column selection.}
\label{sec:exp-colpick}
Before routing to a specialist, \textsc{MetaJoin} must first identify the candidate join columns. We evaluate the input-only certificate using top-$1$ column-pair accuracy and compare it with six individual signals based on exact overlap, token Jaccard, character Jaccard, edit similarity, embedding cosine, and direct \textsc{idfcos} similarity. This decision is required only for the multi-column \textsc{Auto-Join} tables. The remaining single-column benchmarks expose one candidate pair, while the entity-matching benchmarks operate on serialized records. On \textsc{Auto-Join}, transformations weaken surface agreement between the true join columns and cause individual signals to favor lexically similar distractors. No individual signal exceeds $0.58$ top-$1$ accuracy, and both embedding cosine and direct \textsc{idfcos} reach only $0.39$. Combining these signals in the certificate raises top-$1$ accuracy to $0.94$ (Figure~\ref{fig:standalone}a).

\paragraph{Transformation specialist and reciprocal coverage.}
We evaluate \textsc{MNN Join} independently of the controller on \textsc{Auto-Join}, \textsc{FlashFill}, \textsc{SyGuS}, and \textsc{PROSE}, which our workload characterization identifies as transformation-dominant. \textsc{MNN Join} reaches $0.910$ mean row-$F_1$, compared with $0.782$ for \textsc{DeepJoin}~\cite{dong2022deepjoin} and $0.780$ for \textsc{DTT}. Its advantage comes primarily from the reciprocal-coverage objective rather than the underlying string similarity metric. On \textsc{SyGuS}'s \texttt{get-last-name-from-name}, for example, source values such as \texttt{Park Kim} must align with targets such as \texttt{Kim}. Whole-value similarity is ambiguous because surnames can appear in either token position and multiple full names can share the same surname. \textsc{DeepJoin} reaches only $0.40$, while \textsc{DTT} and \textsc{Lotus} both score $0.00$. \textsc{MNN Join} instead selects the final-token projection because it yields the highest reciprocal mutual-nearest-neighbor coverage, producing a perfect join with $F_1=1.0$. Figure~\ref{fig:standalone}b isolates the effect of this objective. Holding the search procedure and similarity function fixed, replacing reciprocal coverage with raw \textsc{idfcos} similarity reduces row-$F_1$ from $0.906$ to $0.890$. Reciprocal coverage wins on $113$ datasets and loses on $44$, as it penalizes the many-to-one collapses that raw similarity can favor. By contrast, changing the underlying similarity function has little effect. With the search fixed, \textsc{idfcos}, IDF $q$-gram, and Ratcliff--Obershelp differ by less than $0.03$ row-$F_1$ (Figure~\ref{fig:standalone}c). We use \textsc{idfcos} for its cost--accuracy tradeoff under the label-free protocol and its efficient implementation as a single vectorized sparse matrix multiplication.

\subsubsection{Adaptive Routing, Validation, and Repair}
\label{sec:ablations}

\begin{figure}[t]
\centering
\includegraphics[width=\columnwidth]{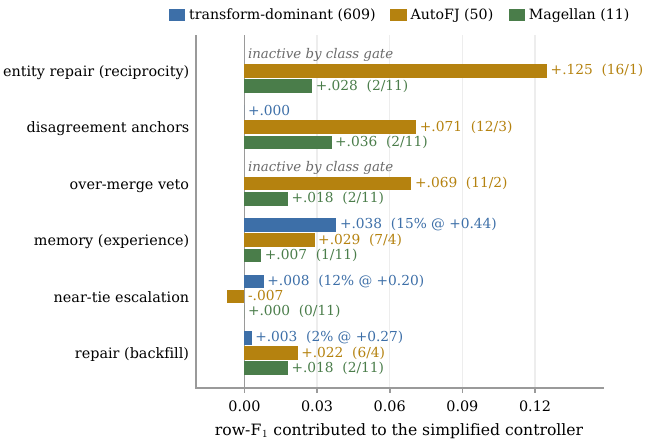}
\caption{Controller mechanism ablations over three seeds. Blue reports the four transformation-dominant benchmarks ($609$ datasets), with affected share and median gain among affected datasets. Gold reports \textsc{AutoFJ} ($50$ tables), with help/hurt counts at $|\Delta|>0.05$. Green reports \textsc{Magellan} ($11$ tables), with affected-table counts at the same threshold.}
\label{fig:ctrlabl}
\end{figure}

\begingroup\emergencystretch=1.5em
Figure~\ref{fig:ctrlabl} removes each controller mechanism independently while holding the rest of the system fixed. The three workload slices exhibit distinct activation patterns. On the transformation-dominant suites, a transformation result carries an exact program witness, so repair is rarely needed. On \textsc{AutoFJ}, the entity witness is sampled and noisy, making output-side repair more important. On \textsc{Magellan}, record structure usually determines the route directly, leaving most repair mechanisms largely inactive. Near-tie escalation changes none of the eleven tables, and every other mechanism changes at most two by more than $0.05$.

\par\endgroup

\paragraph{Output-side validation and repair.}
\begingroup\emergencystretch=1.5em
Output evidence determines whether the initial route should be retained or revised. On the four transformation-dominant benchmarks, the transformation witness is exact, since a pair either satisfies the discovered program or does not, so most joins require little repair after execution. Near-tie escalation compares the primary specialist with its backup when their expected quality is close and contributes $+0.008$ macro row-$F_1$. Its gains concentrate in \textsc{PROSE}, where it improves $12\%$ of datasets by a median of $+0.20$. On \texttt{DateTime.000102} (\S\ref{sec:end2end}), for example, the initial choice between two transformation specialists remains ambiguous, and escalation selects \textsc{DTT}'s join at $0.921$ rather than \textsc{MNN Join}'s $0.059$. Judged backfill addresses a different failure mode by recovering pairs omitted from an otherwise valid partial join and contributes a further $+0.003$. Disagreement-first judging is not triggered on these benchmarks because conflicts between transformation outputs are resolved by method-specific matching evidence, such as overlap between transformed values and candidate targets, rather than by a separate pair judge. The pattern changes on \textsc{AutoFJ}, where the output witness relies on sampled pair judgments and is therefore less decisive. Reciprocity-based entity escalation is the largest contributor at $+0.125\pm0.036$ row-$F_1$. On \texttt{SoccerTournament}, low reciprocal coverage triggers a second entity-matching pass and improves row-$F_1$ by $0.74$ in every seed. The cross-specialist over-merge veto adds $+0.069\pm0.052$, while disagreement-first judging adds $+0.071\pm0.032$ by concentrating pair verification on matches where specialists disagree. On \texttt{Enzyme}, this targeted judging improves row-$F_1$ by about $0.49$ across seeds. Judged backfill contributes another $+0.022\pm0.013$, while near-tie escalation is nearly neutral at $-0.007$. \textsc{Magellan} shows a third pattern, where input structure often makes the initial route decisive before output-side repair is needed. On \texttt{Fodors-Zagats}, record length, field count, and delimiter structure select entity matching without a controller-model call and yield $0.866$ row-$F_1$. The same evidence improves another table from $0.28$ to $0.87$. As a result, near-tie escalation changes none of the eleven tables, and every other mechanism changes at most two by more than $0.05$. The exceptions concentrate in noisier joins. Removing disagreement evidence or entity-side repair lowers row-$F_1$ by about $0.30$ on the two \textsc{Walmart-Amazon} variants in two of three seeds. Together, these results show that MetaJoin applies repair selectively according to the uncertainty remaining after the initial route. Transformation outputs are often supported by stronger method-specific matching evidence, while sampled entity evidence benefits more from escalation and cross-specialist checks. When input structure already makes the route clear, as on most \textsc{Magellan} tables, post-execution repair has little effect.
\par\endgroup


\paragraph{Experience reuse.}
The experience prior reuses evidence from earlier joins to inform later selections. Across three random stream orderings, pooled row-$F_1$ on the four transformation-dominant benchmarks remains $0.9005\pm0.0008$, compared with $0.9006$ under the fixed order, showing little sensitivity to dataset order. In a continuous pass with one shared store, accuracy remains $0.901$ while controller calls fall from $3.9$ per dataset on the initial cold benchmark to $1.9$--$2.5$ as evidence accumulates. Removing the prior changes only $15\%$ of outputs, but on these datasets it contributes a median gain of $0.436$ row-$F_1$, including $+0.10$ on \textsc{PROSE} and $+0.05$ on \textsc{SyGuS}. The gains can be substantial when prior experience corrects an otherwise poor route. For example, \texttt{State.000003} and \texttt{Phone.000007} both fail without the store but reach $1.0$ row-$F_1$ when a similar earlier route is reused. On \textsc{AutoFJ}, the prior contributes $+0.029\pm0.029$, while on \textsc{Magellan} its only notable effect is a $+0.08$ gain on \texttt{Structured\_iTunes-Amazon}. Experience reuse therefore reduces repeated controller effort while selectively improving joins for which prior routing evidence is informative.

\paragraph{Portfolio sensitivity.}
Finally, we separate the benefit of adaptive selection from the capabilities available in the portfolio. With the full portfolio, \textsc{MetaJoin} reaches $0.915$ row-$F_1$ averaged over the four transformation-dominant benchmarks, compared with $0.910$ for the strongest standalone specialist. Removing \textsc{MNN Join} lowers performance to $0.874$, but \textsc{MetaJoin} still exceeds the strongest remaining specialist at $0.791$. The removal affects $23\%$ of datasets, with a median loss of $0.154$ among those affected, while the remaining datasets are successfully rerouted. For example, \texttt{sharif username to email} and \texttt{beatles songs} retain their original scores because \textsc{DTT} provides overlapping capability after \textsc{MNN Join} is removed. In contrast, \texttt{Address.000013} falls from $1.0$ to $0.13$ because its extraction pattern is recovered only by \textsc{MNN Join}'s transformation search. Removing both \textsc{MNN Join} and \textsc{DTT} affects $43\%$ of datasets and lowers performance to $0.776$, slightly below the strongest surviving method at $0.791$. These results show that adaptive selection is robust to portfolio changes when specialists provide overlapping capabilities, but cannot recover a matching capability absent from the portfolio.

\begin{figure}[t]
\centering
\includegraphics[width=\columnwidth]{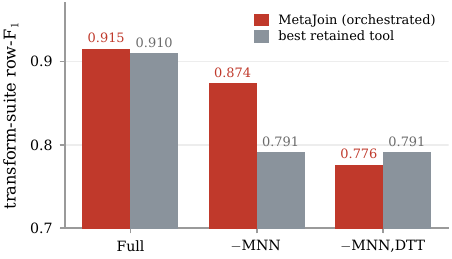}
\caption{Portfolio sensitivity. \textsc{MetaJoin} is compared with the strongest standalone method retained after removing specialists from the portfolio.}
\label{fig:toollib}
\end{figure}

\subsection{RQ3: Efficiency, Scalability, and Robustness}
\label{sec:deployment}

Adaptive join selection should improve accuracy without sacrificing the efficiency gained from selective execution. We therefore measure token and model-call cost, end-to-end runtime, scalability with table size, and sensitivity to the controller backbone.

\subsubsection{Execution Cost and Scalability}
\label{sec:efficiency}

\paragraph{Accuracy--cost tradeoff.}
In table~\ref{tab:results}, embedding and program-synthesis methods, including \textsc{MNN Join}, \textsc{Semantic Join}, \textsc{DeepJoin}, \textsc{GXJoin}~\cite{omidvartehrani2024gxjoin}, and \textsc{Auto-Join-PBE}~\cite{zhu2017auto}, require no model calls. The two PBE baselines use the same $25$ transformation operators as \textsc{MNN Join} (Table~\ref{tab:operators}), ensuring that differences arise from the search strategy rather than the operator set. In contrast, per-pair and per-value LLM methods such as \textsc{Lotus}, the entity matcher, and \textsc{FM-Wrangle} spend a median of $7$--$31$k tokens per join and can exceed $10^{6}$ tokens on large entity tables as cost grows with the candidate space. \textsc{MetaJoin} has a median cost of about $1$k tokens per dataset because most joins require only controller evidence calls and a token-free specialist. Its mean rises to $177$k because the entity matcher dominates cost on the subset of joins that require it. This remains below run-all-select at $194$k, while achieving substantially higher accuracy. MetaJoin therefore allocates expensive matching selectively rather than invoking it across the full workload.

\begin{figure*}[t]
\centering
\includegraphics[width=\textwidth]{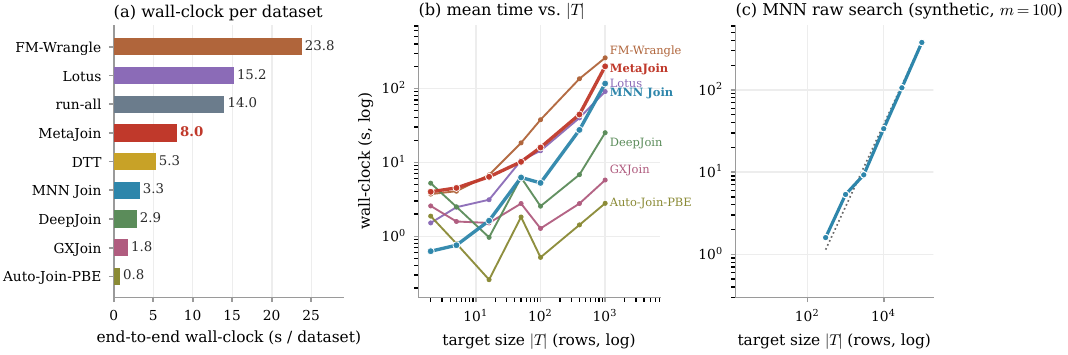}
\caption{Cost and scalability. (a) End-to-end wall-clock time on the four transformation-dominant benchmarks. (b) Transformation-search time versus target size on a log--log scale. (c) \textsc{MNN Join}'s raw reciprocal search on synthetic tables.}
\label{fig:cost-scaling}
\end{figure*}

\paragraph{Wall-clock cost and scalability.}
Figure~\ref{fig:cost-scaling}a compares end-to-end runtime on the four transformation-dominant benchmarks. \textsc{MetaJoin} takes about $8$ seconds per dataset on its common \textsc{MNN Join}+\textsc{DTT} path, compared with $14.0$ seconds for executing the full portfolio. It is also $2$--$3\times$ faster than the per-pair LLM methods \textsc{Lotus} ($15.2$ seconds) and \textsc{FM-Wrangle} ($23.8$ seconds). Token-free baselines are faster on average at $0.8$--$2.9$ seconds, but achieve substantially lower accuracy. On \textsc{Auto-Join}, where transformation search is required for most datasets, \textsc{MNN Join} averages $3.3$ seconds compared with $4.3$ for \textsc{Auto-Join-PBE}, $4.8$ for \textsc{GXJoin}, and $16.9$ for \textsc{DeepJoin}. Figure~\ref{fig:cost-scaling}b further shows how these costs change with table size on the observed workload. Per-pair LLM methods grow from a few seconds to $90$--$260$ seconds as the target approaches $1{,}000$ rows, while the token-free transformation methods remain below $10$ seconds and \textsc{MetaJoin} follows the cost of the specialists selected at each scale. To examine scaling beyond the benchmark range, we then run a controlled synthetic sweep because the real tables contain at most about $2{,}000$ rows. At fixed source size and target sizes up to $10^5$, \textsc{MNN Join}'s raw sparse-matrix search grows from $1.6$ to $378$ seconds, approximately linearly with target size and consistent with its $O(mn)$ probe (\S\ref{sec:mnnjoin}) (Figure~\ref{fig:cost-scaling}c).  Together, the three panels show that \textsc{MetaJoin} avoids the steep runtime growth of per-pair LLM methods through selective execution, while \textsc{MNN Join} scales predictably to substantially larger tables.

\subsubsection{Backbone Robustness}
\label{sec:generality}

\textsc{MetaJoin} uses an LLM for several selection and verification decisions, so we test whether its performance depends on a particular backbone. We replace the reference \textsc{Qwen3-8B}-Instruct model with the base releases of \textsc{Llama-3.2-3B}, \textsc{Llama-3.1-8B}, \textsc{Qwen3-4B}, and \textsc{Qwen3-14B}. For each backbone $M$, we replace the model at every swappable point, including the catalog selector, content probe, pair verifier, and the example-guided transformation specialist regenerated with $M$. We keep the off-the-shelf entity matcher fixed because its fine-tuned model defines the specialist itself.

\begin{figure}[t]
\centering
\includegraphics[width=\columnwidth]{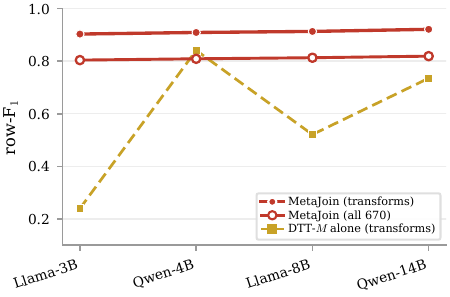}
\caption{Full-system backbone swap. Each fully-$M$ configuration replaces every swappable LLM component with backbone $M$, including the LLM-based transformation specialist.}
\label{fig:generality}
\end{figure}

\paragraph{Accuracy is substantially less sensitive to the backbone.}
Figure~\ref{fig:generality} compares the resulting fully-$M$ systems with LLM-based join baselines evaluated under the same backbone swaps. The baselines vary sharply with model choice. LLM semantic join spans $0.26$ row-$F_1$, the example-guided transformation baseline spans $0.60$, and the per-pair foundation-model wrangler varies by more than $0.4$. The transformation baseline is also non-monotonic in model size, with its $4$B configuration reaching $0.841$ compared with $0.735$ at $14$B. In contrast, \textsc{MetaJoin} varies by only $0.015$ macro row-$F_1$ across the $3$B--$14$B backbones. This stability persists even when the LLM-based transformation specialist is degraded together with the controller. Under the weakest backbone, that specialist falls to $0.240$ row-$F_1$ on the transformation-dominant benchmarks, while the full \textsc{MetaJoin} system remains at $0.903$. Stronger backbones primarily reduce decision cost rather than change final accuracy. Across all $670$ datasets, macro row-$F_1$ ranges from $0.804$ at $3$B to $0.819$ at $14$B, while model calls fall from $6.7$ to $3.1$ per task.

\paragraph{Validation and repair absorb backbone variation.}
The backbones do not make the same initial selections. On the transformation-dominant workload, the $3$B model selects a different specialist set from the reference on $58\%$ of datasets, while the $14$B model differs on $32\%$. If each model's first routed specialist is committed without repair, row-$F_1$ ranges from $0.788$ for the $3$B backbone to $0.915$ for the $14$B backbone, a gap of $0.127$. Validation and repair compress this gap to $0.018$. The recovery is largest for the weakest model, adding $0.115$ row-$F_1$ at $3$B compared with $0.006$ at $14$B. The same effect appears when the transformation specialist itself degrades. Even when its standalone row-$F_1$ falls below $0.25$, weak transformation outputs are rejected by method-specific output evidence and alternative specialists absorb the unresolved joins. Entity-matching performance is nearly flat across backbones at $0.63$--$0.64$ because the entity specialist is held fixed. These results show that backbone quality primarily affects the quality and cost of the initial selection, while output validation and repair reduce its effect on the final join.

%% file: figures/master_table.tex
\begin{table*}[t]
\centering
\definecolor{ourred}{HTML}{C0392B}
\caption{Label-free row-$F_1$ (mean$\pm$std, three seeds; shaded by value), the benchmark macro mean, and mean LLM tokens per task ($a$: controller, $b$: specialists; $0$: no model calls). \textbf{Bold}: best per column.}
\label{tab:results}
\setlength{\tabcolsep}{3pt}\small
\begin{tabular}{@{}l cccccc c r@{}}
\toprule
Method & Auto-Join & FlashFill & SyGuS & PROSE & AutoFJ & Magellan & Macro & Tokens\,($a{+}b$)\\
\midrule
\multicolumn{9}{@{}l}{\emph{External baselines}}\\
DeepJoin & \cellcolor{ourred!40}.81{\tiny$\pm$.000} & \cellcolor{ourred!42}.83{\tiny$\pm$.000} & \cellcolor{ourred!40}.80{\tiny$\pm$.000} & \cellcolor{ourred!34}.69{\tiny$\pm$.000} & \cellcolor{ourred!8}.17{\tiny$\pm$.000} & \cellcolor{ourred!0}.00{\tiny$\pm$.000} & \cellcolor{ourred!28}.55\,{\scriptsize$\pm$.000} & 0\\
Lotus & \cellcolor{ourred!34}.69{\tiny$\pm$.001} & \cellcolor{ourred!42}.83{\tiny$\pm$.000} & \cellcolor{ourred!40}.79{\tiny$\pm$.001} & \cellcolor{ourred!31}.62{\tiny$\pm$.001} & \cellcolor{ourred!24}.49{\tiny$\pm$.001} & \cellcolor{ourred!25}.50{\tiny$\pm$.003} & \cellcolor{ourred!33}.65\,{\scriptsize$\pm$.000} & 7k\\
FM-Wrangle & \cellcolor{ourred!17}.34{\tiny$\pm$.000} & \cellcolor{ourred!10}.21{\tiny$\pm$.000} & \cellcolor{ourred!6}.11{\tiny$\pm$.000} & \cellcolor{ourred!8}.16{\tiny$\pm$.000} & \cellcolor{ourred!26}.52{\tiny$\pm$.000} & \cellcolor{ourred!25}.50{\tiny$\pm$.000} & \cellcolor{ourred!15}.31\,{\scriptsize$\pm$.000} & 31k\\
GXJoin & \cellcolor{ourred!12}.23{\tiny$\pm$.000} & \cellcolor{ourred!11}.22{\tiny$\pm$.000} & \cellcolor{ourred!17}.34{\tiny$\pm$.000} & \cellcolor{ourred!12}.25{\tiny$\pm$.000} & \cellcolor{ourred!6}.11{\tiny$\pm$.000} & \cellcolor{ourred!0}.00{\tiny$\pm$.000} & \cellcolor{ourred!10}.19\,{\scriptsize$\pm$.000} & 0\\
Auto-Join-PBE & \cellcolor{ourred!18}.36{\tiny$\pm$.000} & \cellcolor{ourred!4}.07{\tiny$\pm$.000} & \cellcolor{ourred!2}.04{\tiny$\pm$.000} & \cellcolor{ourred!6}.12{\tiny$\pm$.000} & \cellcolor{ourred!6}.12{\tiny$\pm$.000} & \cellcolor{ourred!0}.00{\tiny$\pm$.000} & \cellcolor{ourred!6}.12\,{\scriptsize$\pm$.000} & 0\\
\midrule
\multicolumn{9}{@{}l}{\emph{Specialists}}\\
MNN Join & \cellcolor{ourred!43}.86{\tiny$\pm$.001} & \cellcolor{ourred!49}\textbf{.98}{\tiny$\pm$.000} & \cellcolor{ourred!48}\textbf{.96}{\tiny$\pm$.000} & \cellcolor{ourred!42}.84{\tiny$\pm$.002} & \cellcolor{ourred!18}.36{\tiny$\pm$.012} & \cellcolor{ourred!0}.00{\tiny$\pm$.000} & \cellcolor{ourred!33}.67\,{\scriptsize$\pm$.002} & 0\\
DTT & \cellcolor{ourred!40}.79{\tiny$\pm$.005} & \cellcolor{ourred!44}.88{\tiny$\pm$.012} & \cellcolor{ourred!40}.81{\tiny$\pm$.014} & \cellcolor{ourred!32}.64{\tiny$\pm$.004} & \cellcolor{ourred!6}.11{\tiny$\pm$.001} & \cellcolor{ourred!0}.00{\tiny$\pm$.000} & \cellcolor{ourred!27}.54\,{\scriptsize$\pm$.003} & 1k\\
Semantic Join & \cellcolor{ourred!40}.81{\tiny$\pm$.000} & \cellcolor{ourred!42}.83{\tiny$\pm$.000} & \cellcolor{ourred!40}.81{\tiny$\pm$.000} & \cellcolor{ourred!36}.72{\tiny$\pm$.000} & \cellcolor{ourred!8}.17{\tiny$\pm$.000} & \cellcolor{ourred!20}.39{\tiny$\pm$.000} & \cellcolor{ourred!31}.62\,{\scriptsize$\pm$.000} & 0\\
Equality & \cellcolor{ourred!2}.05{\tiny$\pm$.000} & \cellcolor{ourred!2}.04{\tiny$\pm$.000} & \cellcolor{ourred!2}.04{\tiny$\pm$.000} & \cellcolor{ourred!4}.09{\tiny$\pm$.000} & \cellcolor{ourred!6}.12{\tiny$\pm$.000} & \cellcolor{ourred!0}.00{\tiny$\pm$.000} & \cellcolor{ourred!3}.06\,{\scriptsize$\pm$.000} & 0\\
Entity matcher & \cellcolor{ourred!32}.65{\tiny$\pm$.000} & \cellcolor{ourred!26}.53{\tiny$\pm$.000} & \cellcolor{ourred!24}.47{\tiny$\pm$.000} & \cellcolor{ourred!19}.38{\tiny$\pm$.000} & \cellcolor{ourred!33}\textbf{.66}{\tiny$\pm$.000} & \cellcolor{ourred!28}\textbf{.57}{\tiny$\pm$.000} & \cellcolor{ourred!27}.54\,{\scriptsize$\pm$.000} & 192k\\
\midrule
\multicolumn{9}{@{}l}{\emph{Label-free policies (same specialist library + \textsc{Qwen3-8B-Instruct} decision model)}}\\
LLM router & \cellcolor{ourred!43}.86{\tiny$\pm$.001} & \cellcolor{ourred!49}\textbf{.98}{\tiny$\pm$.000} & \cellcolor{ourred!48}\textbf{.96}{\tiny$\pm$.000} & \cellcolor{ourred!42}.83{\tiny$\pm$.001} & \cellcolor{ourred!8}.15{\tiny$\pm$.000} & \cellcolor{ourred!14}.29{\tiny$\pm$.021} & \cellcolor{ourred!34}.68\,{\scriptsize$\pm$.003} & 0.6k\,{\scriptsize(0.5{+}0.1)}\\
run-all-select & \cellcolor{ourred!40}.80{\tiny$\pm$.007} & \cellcolor{ourred!43}.86{\tiny$\pm$.007} & \cellcolor{ourred!44}.87{\tiny$\pm$.021} & \cellcolor{ourred!38}.75{\tiny$\pm$.004} & \cellcolor{ourred!12}.24{\tiny$\pm$.013} & \cellcolor{ourred!22}.44{\tiny$\pm$.000} & \cellcolor{ourred!33}.66\,{\scriptsize$\pm$.004} & 194k\,{\scriptsize(0.6{+}193)}\\
\textbf{MetaJoin} & \cellcolor{ourred!44}\textbf{.88}{\tiny$\pm$.004} & \cellcolor{ourred!48}.97{\tiny$\pm$.001} & \cellcolor{ourred!48}\textbf{.96}{\tiny$\pm$.009} & \cellcolor{ourred!44}\textbf{.87}{\tiny$\pm$.005} & \cellcolor{ourred!32}.65{\tiny$\pm$.005} & \cellcolor{ourred!28}\textbf{.57}{\tiny$\pm$.053} & \cellcolor{ourred!41}\textbf{.82}\,{\scriptsize$\pm$.001} & 177k\,{\scriptsize(1.1{+}176)}\\
\bottomrule
\end{tabular}
\end{table*}

%% file: discussion.tex

\section{Conclusion}
\label{sec:conclusion}

We presented \textsc{MetaJoin}, a label-free framework for adaptive join selection across heterogeneous tables. Rather than treating specialist selection as a one-shot routing decision, \textsc{MetaJoin} uses input evidence to choose an initial specialist and specialist-specific output witnesses to validate, compare, or revise that choice. Its transformation specialist, \textsc{MNN Join}, recovers transformations directly from unaligned columns through reciprocal alignment without requiring labeled examples. Across six benchmarks, \textsc{MetaJoin} slightly exceeds a benchmark-level oracle that is given the best standalone specialist for each benchmark and outperforms label-free policies built from the same specialist library by more than $0.13$ macro row-$F_1$. Its accuracy also remains stable as the controller backbone varies from $3$B to $14$B parameters because validation and repair absorb many errors in the initial route. Together, these results show that reliable join selection depends not only on choosing a suitable matching method from the input, but also on using evidence produced during matching to determine whether that choice should be retained or revised.